# Magnetic Skyrmions for Unconventional Computing

Sai Li[1,2], Wang Kang[1], Xichao Zhang[3], Tianxiao Nie[1], Yan Zhou[3],

Kang L. Wang[4] and Weisheng Zhao[1]

[1]School of Integrated Circuit Science and Engineering, Beijing Advanced Innovation Center for Big Data and Brain Computing, Beihang University, Beijing, 100191, China.

[2]Shenyuan Honors College, Beihang University, Beijing, 100191, China.

[3]School of Science and Engineering, The Chinese University of Hong Kong, Shenzhen, 518172, China.

[4]Department of Electrical and Computer Engineering, University of California, Los Angeles, CA, 90095, USA

*Corresponding E-mails: wang.kang@buaa.edu.cn; weisheng.zhao@buaa.edu.cn

## Abstract

Improvements in computing performance have significantly slowed down over the past few years owing to the intrinsic limitations of computing hardware. However, the demand for data computing has increased exponentially. To solve this problem, tremendous attention has been focused on the continuous scaling of the Moore's Law as well as the advanced non-von Neumann computing architecture. A rich variety of unconventional computing paradigms has been raised with the rapid development of nanoscale devices. Magnetic skyrmions, spin swirling quasiparticles, have been endowed with great expectations for unconventional computing due to their potential as the smallest information carriers by exploiting their physics and dynamics. In this paper, we provide an overview of the recent progress of skyrmion-based unconventional computing from a joint device-application perspective. This paper aims to build up a panoramic picture, analyze the remaining challenges, and most importantly to shed light on the outlook of skyrmion based unconventional computing for interdisciplinary researchers.

## 1 Introduction

### 1.1 What and why unconventional computing

With megatrends of cloud computing, big data and artificial intelligence (AI), the age of Internet in the decade is driving a data explosion. It has revolutionized the way of data generation, storage and processing. Along with benefits from the flashy apps, rapid ascent of social media and prompt mobile payment etc., people may inadvertently create 2.5 quintillion bytes of data per day [1]. The resulting demands on data storage and processing increase exponentially [2]. In the past, algorithm/software is usually the center of attention when industry mentions the advance of information processing. Recently, however, hardware that underlies all these progresses has earned worldwide concern [3].

The Von Neumann computing architecture, for most computers available today, even since its invention, however, gradually fails to meet the requirements of real-time high-performance data processing with the explosive growth of data volume [4]. Although from 1985to 2010s, the architectural innovations, e.g. using multiple levels of on-chip cache memories, have gained up to 100-fold computing performance, the so-called von Neumann bottleneck—"memory wall" still exists. This age-old problem lies in the physical separation between the processing unit and the memory unit, leading to the inadequate performance caused by the constrained bandwidth of the data bus communicating the two [5]. A good fit for current semiconductor technology to solve this challenge is called in-memory computing, which processes data in-situ within the memory unit where data are stored [6], significantly reducing the data transportation and breaking the memory wall. Unfortunately, the hardware/software co-design faces big challenges of achieving high-performance in-memory computing architecture [7]. One of the biggest challenges is to search for a good memory medium that supports the integration of data storage and computing in a single die. Beyond in-memory computing, a rich variety of unconventional computing technologies have emerged tending to solve the Von Neumann bottleneck (Fig. 1). How to define it, in contrast to conventional Von Neumann design, is still an open question, which breaks boundaries in thinking, acting and processing [8]. In this context, the exploration of different materials, devices and architectures in unconventional computing has provoked an emerging research interest in academia and industry.

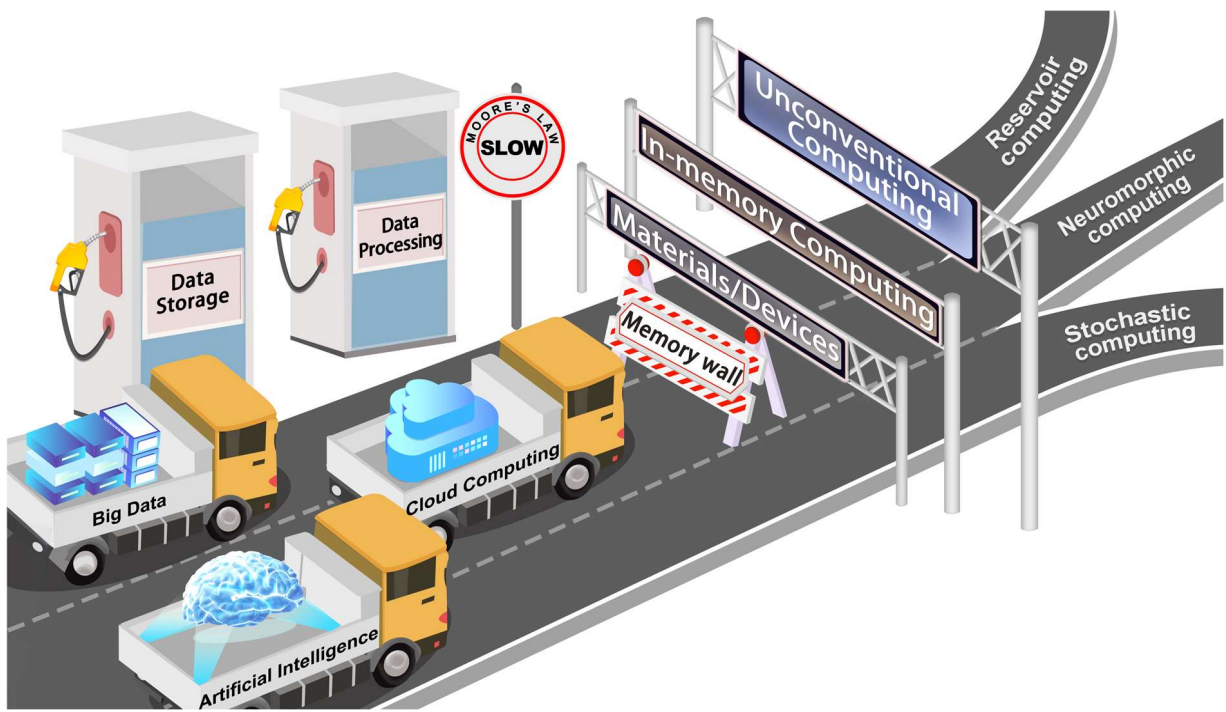

Fig.1. The roadmap towards future unconventional computing solutions [9].

## 1.2 Magnetic skyrmions as new information carriers

It has been widely recognized that the method of relying solely on size reduction to further improve computing performance is no longer applicable because of the physical limit of silicon transistors [10]. Intensive effort has been underway to search for next-generation information carriers to keep continuously increasing the performance while maintaining low power and high reliability. One of the most promising candidates is spintronics, where the intrinsic spin freedom of electron in addition to charge and its associated magnetic moment are manipulated to avoid energy penalty [11]. Meanwhile, various emerging nonvolatile technologies have been developed to apply unconventional technologies [12], showing different advantages and challenges. For instance, resistive RAM (RRAM) has low energy cost and high density but is limited by endurance and compatibility with CMOS [13]. Phase change memory (PCM) owns the efficient conductance tunability while is challenging by inherent material issues of resistance drift and phase relaxation [5]. Therefore, featuring high endurance, high speed, CMOS compatibility and low-power makes spintronic devices stand out. Accompanied by the commercial production of spin-transfer-torque magnetoresistive random access memory (STT-MRAM) and giant magnetoresistive (GMR)/tunneling magnetoresistive (TMR) sensors, there has been substantial progress in developing spintronic devices for unconventional computing applications, such as the neuromorphic computing [14]. Several proposals have been reported to design artificial synapses and spiking neurons by exploring magnetic tunnel junction (MTJ) [15], spin valve [16], spin torque nano-oscillator (STNO) [17] and domain wall motion (DWM) [18] for neuromorphic computing.

Recent advances in spintronic devices are primarily motivated by two technologies [19], i.e. controlling magnets via spin torques [20,21] and the discovery of topological spin textures [22]. Electric-field/current induced magnetization dynamics has greatly promoted the development of spintronic memories. On the other hand, topological spin textures, e.g. magnetic skyrmion, have been of great interest as new information carrier owing to their non-trivial topology, nanoscale size, defect-tolerance, low driving current properties. Compared to other magnetic devices, skyrmion-based devices could be smaller, more stable (less sensitive to defects in materials), more efficiently adjustable (easier to change the skyrmions numbers and utilize ample interactions). The advantages of magnetic skyrmion could enable new possibilities/functionalities that may be inaccessible to conventional semiconductor technology, paving the way towards unconventional computing hardware revolution.

The application of skyrmion was firstly introduced in designing racetrack memory, a high-density memory technology proposed by S. Parkin in 2008 [23]. Unfortunately, owing to the high demand for write/read accuracy and reliability, which are challenged by material impurities/defects [24], as well as process/temperature variations [25], magnetic skyrmion based racetrack memory remains lots of challenges even with great advances in the past ten years. In this context, unconventional computing, can be an alternative by utilizing the variability, uncertainty and dynamic behaviors of magnetic skyrmion. To date, a variety of publications have summarized the key progress and future research directions of magnetic skyrmion in materials and physics [26–28], while a systematic review of magnetic skyrmion in unconventional computing has not yet been reported, which, we believe, is necessary for researchers in the interdisciplinary research area. This review focuses on the research progress and physical behaviors of magnetic skyrmion-based unconventional computing, where, in particular, device concepts and related circuits/algorithms will lend perspectives for future magnetic skyrmion based electronics.

## 2  Dynamic behaviors of magnetic skyrmions

This section briefly introduces the dynamic behaviors of magnetic skyrmion before we review its application in unconventional computing. The name of skyrmion comes from the physicist Tony Skyrme [29], who proposed the topological model of the nucleon. Then Bogdanov et al. predicted it as topological soliton theoretically in condensed matter systems [30]. Nowadays, studies of skyrmion have focused on the spintronics filed since the first observation of skyrmion is in magnetic materials (refer to magnetic skyrmion) with non-centrosymmetric B20 lattices through a small angle neutron scattering (SANS) experiment [31]. Hereafter, we abbreviate "magnetic skyrmion" as "skyrmion" for convenience.

Physically, skyrmion is a new type of magnetic winding configuration, mostly the chirality of which originates from antisymmetric Dzyaloshinskii–Moriya interaction (DMI) existing in systems lacking inversion symmetry. Different symmetries of the spin-orbit coupling (SOC) interaction could be achieved by the material composition (e.g. crystal lattice with Dresselhaus SOC or interface-engineered thin films with Rashba SOC) [32]. Here, the energy density of the interfacial DMI provided by the DMI effective field is considered as

$$\varepsilon_{DM} = m_z(\nabla \cdot \vec{m}) - (\vec{m} \cdot \nabla)m_z = -\frac{1}{2}\vec{M} \cdot \vec{B}_{DM} \quad (1)$$

Correspondingly, there are two typical types of skyrmions with different vorticities and helicities: Bloch-type skyrmion and Néel-type skyrmion (Fig. 2a,b) [33,34]. From the view of implementable applications, skyrmions in thin-films are more competitive than those in bulk compounds for the following reasons [27,28]: (1) room temperature stability; (2) zero magnetic field requirement; (3) material growth compatibility with Complementary Metal Oxide Semiconductor (CMOS) technologies.

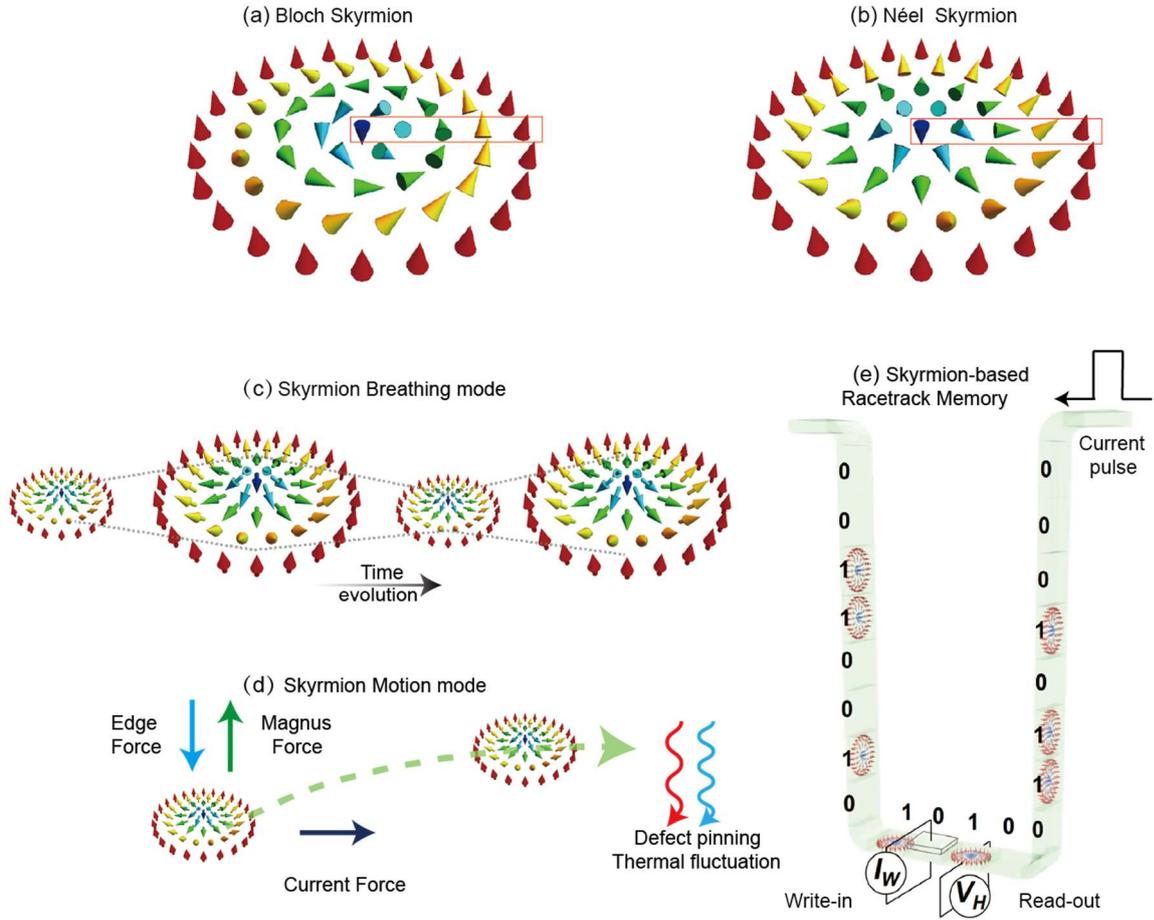

Fig.2. Two basic dynamic modes of magnetic skyrmions. (a) Bloch skyrmion. (b) Néel skyrmion. (c) Skyrmion breathing mode with skyrmion size oscillation. (d) Skyrmion motion mode motivated by edge force, magnus force and current force. (e) The skyrmion-based racetrack memory[35]. (Copyright 2015, Nature Publishing Group.)

To date, among various methods to manipulate individual skyrmions [36,37], the most promising approach is the electrical method, which hosts potential for high-density and low-power devices. Here, we use the Thiele approach to understand the dynamic process of skyrmion with the assumption that the skyrmion is a rigid object:

$$\boldsymbol{G} \times \frac{d\boldsymbol{X}}{dt} + \frac{dU}{d\boldsymbol{X}} + \boldsymbol{\mathcal{D}}\frac{d\boldsymbol{X}}{dt} = \boldsymbol{F}_{tot} \quad (2)$$

where we take the gyromagnetic coupling vector $\boldsymbol{G}$, $\frac{d\boldsymbol{X}}{dt}$ is the velocity of skyrmion, $U$ describes the boundary potential acting on the skyrmion, $\boldsymbol{\mathcal{D}}$ is the dissipative force tensor, and $\boldsymbol{F}_{tot}$ is mainly given by spin current force (STT or spin orbit torque (SOT)). Because of the first term in the left side of the equation, i.e. the topological Magnus force, skyrmion would deviate from the driving current direction perpendicularly and even cause its annihilation, leading to the skyrmion Hall effect (SkHE). In addition, diffusive skyrmion motion at room temperature and skyrmion-skyrmion interaction will lead to irregular velocity of skyrmions and the ubiquitous impurity effects will pin skyrmions unexpectedly. All of these make the linear shift of skyrmions practically impossible, thereby skyrmion-based information bits would not be read in a traditional manner. Moreover, most skyrmions are equivalent to metastable states, where the unpredictable deletion and nucleation could happen by the noise currents or fields [38]. Therefore, to study the dynamics of skyrmion exploited in unconventional applications is necessary and can be two-fold. Firstly, the breathing-like behaviors of skyrmion perform as regular size variation (Fig. 2c) and the localized magnetization fluctuates topologically with a certain frequency [39]. Secondly, the skyrmion motion mode could be manipulated by multiple forces (Fig. 2d) and designed to perform different functions suited to nanodevices. Many conceptual proposals of skyrmion-based nanodevices have been published, especially some of which have been realized in experiment recently. The room-temperature stability of skyrmion in ultrathin films [40] makes it increasingly attractive to be an effective candidate for the unconventional computing.

## 3 Skyrmion-based unconventional computing devices

Skyrmion was first proposed to be used in horizontal racetrack memories [41], where information is coded by the presence or absence of a skyrmion in a nanotrack (Fig. 2e). Later, skyrmionic logic gates [21,42] and transistor-like devices [43] were introduced, tending to replace conventional semiconductor digital designs . However, there exist some inevitable challenges hindering the accuracy of such devices in large-scale computing systems, such as natural imperfections/defects in magnetic thin films and thermal fluctuations. On the other hand, skyrmionic unconventional computing devices, which exploit these features, have gained well-reasoned attention gradually. In this section, we will introduce state-of-the-art mainstream skyrmionic unconventional computing devices (see table.1), in terms of microwave devices, neuromorphic computing devices (including artificial synaptic and neuron devices), reservoir computing devices and stochastic computing device.

### 3.1 Skyrmionic microwave devices

STNO, based on an MTJ stucture, is used to generate spontaneous microwave with specific frequencies. Skyrmionic STNO combines the advantages of skyrmion and STNO, and could extend the range of working frequency tremendously while markably reducing the starting oscillation delay [44]. When a single skyrmion is integrated in a STNO, gigahertz oscillation can be activated vigorously once started by a DC spin-polarized current and the amplitude can be rapidly stabilized to a steady value (Fig. 3d). The skyrmion is driven by the STT of the center electrode and then detected by the ones with centrosymmetric distribution at the edge of the disk, of which number and location could be tuned depending on the device efficiency. It should be noted that the working frequency, output power and linewidth are three most essential parameters delivered from the skyrmionic STNO. The total energy of the device varies with almost a linear function corresponding to the square of the skyrmion core area (Fig. 3b). The oscillation frequency is mainly determined by the edge potential, namely the size of the center contact and the nanodisk, instead of spin transfer torque force of the DC current, which could range from 0 Hz to about 1.4 GHz theoretically (Fig. 3a). Particularly, the phase of the output microwave signal could be different as there are different pairs of detection contacts. A strength of this device is that its linewidth (full width at half maximum of the power spectra) is less than 1 MHz theoretically [44], . While for a single skyrmion, two limitations include: (a) the output power is confined and (b) the starting oscillation time is long. It is possible to expand the range of working frequency, enlarge the output power and reduce the starting oscillation time (Fig. 3c), by employing multiple skyrmions, which requires that the number of skyrmions and the detection contacts should be identical to ensure synchronism.

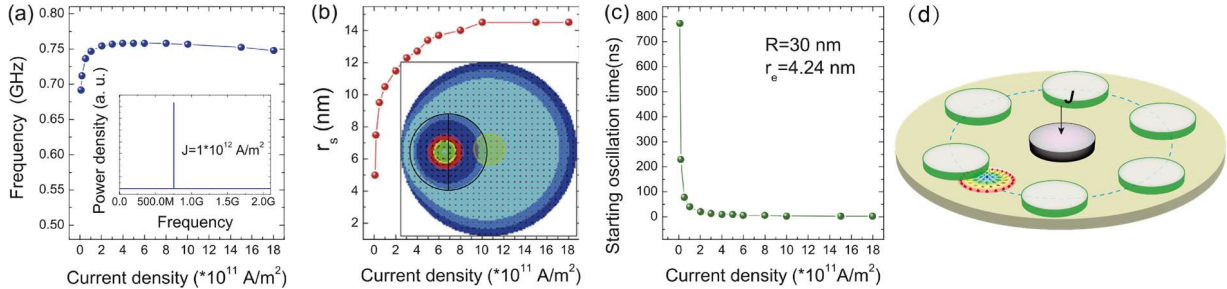

Fig.3. Skyrmionic STNO device. The oscillation frequency (a), oscillation radius (b) and starting oscillation time (c) as a function of the DC current density. The inset in (a) is the FFT power figure. The inset in (b) is the contour plots of *Mz*. (d) Schematic diagram of the device with multiple pairs of detection electrodes. When the skyrmion moves into the area of the detection electrodes, the resistance will be high; otherwise the resistance becomes low [44]. (Copyright 2015, IOP Publishing Group.)

In 2015, Finocchio et al. proposed the use of the dynamical skyrmion response to current through the spin-transfer torque diode (STD) effect, related to ferromagnetic (FM) resonance (FMR) in the basic principle, to detect and harvest microwave passively [45]. The researchers converted the skyrmion size oscillation, rather than skyrmion motion oscillation, to a change of the TMR rendering a detection voltage. The skyrmionic STD device characterizes the excited skyrmion response of a breathing mode in a quasi-linear regime, which creates a resistance oscillation by shaking the skyrmion core diameter. An optimal configuration indicates that the sensitivity of the skyrmionic STD is of the order of 2000V/W, which can also be improved by the voltage-controlled magnetic anisotropy (VCMA) effect, much more than the ones of the unbiased MTJ-based STD with nearly 900 V/W. The idea of skyrmionic microwave devices paves the way for exploring skyrmion from fundamental physics to designing ultra-low-power unconventional electronic devices in microwave area [46–48]. It should be noted that the issue may arise when intergrating multiple skyrmionic STNOs, where the physical interactions among devices would have a nonnegligible impact. The further exploration of the integrated system ( e.g., device interval) is highly needed.

## 3.2 Skyrmionic neuromorphic computing devices

Inspired by the biological brain, neuromorphic computing recaptures considerable attention accompanied by the growing popularity of artificial intelligence (AI) and the advances of emerging nanoscale devices. Conventional computing platforms, e.g. central processing units (CPUs) and graphic processing units (GPUs), cannot satisfy the demand of running large-scale neural network, in terms of power consumption and computing speed, due to the dissimilarities between the neural network model and the hardware. Instead, neuromorphic computing imitates the physical structure of a human brain from computational primitive elements – spiking neurons to their connectivity pattern – synaptic plasticity, establishing spike event-driven communication, thus improving temporal-spatial efficiency of the hardware (Fig. 4). There are two components of the neural elemental functionalities: neuron and synapse, both of which appear sophisticated time-dependent behaviors. It is estimated that more than 100 billion neurons and each neuron maintains roughly 1,000 connective synapses with other ones, some even linking 200,000 synapses. In this context, neuromorphic chips and non-volatile technologies provide different energy-efficient solutions closer to a neurobiological basis by mimicking aspects of the brain's architecture and dynamics [49]. However, from the perspective of a neuromorphic computing device, which fulfills all the following needs, such as nanoscale size, long write/erase endurance, long retention, nanosecond switching speed, and low programming energy, has not been found [9]. The emergence of skyrmion provides a new possibility for designing neuromorphic computing devices to imitate the hallmark functional capabilities of the human brain.

### Skyrmionic synaptic devices

In a nervous system, a synapse acts as a switching medium to pass an electrical or chemical signal from one neuron to another. Therefore, in order to be closer to biological characteristics, the synaptic weights are expected to be real-valued rather than binary, the former of which is easy to realize in software simulation but harder in existing silicon devices. Most semiconductor-based neuromorphic circuits need to integrate many transistors at the cost of energy efficiency and integration density [50–52]. Consequently, novel designs have been coined as an alternative, which can incorporate memristor devices [53–55], phase change memories [56,57], magnetic tunnel junctions [58,59], domain wall magnets [60,61], atomic switches [62,63], and organic resistance switch [32], etc. Although these devices are potential candidates, they have different restrictions on either fundamental form of learning rule for synapse [64] or limited scalability and stability [65]. On the other hand, the information communication in a biological synapse is realized by releasing neurotransmitters. This particle-like substance plays an essential role in propagating spikes in either potentiation or depression modes. Analogously, skyrmions are particle-like spin swirling with topological stability and could be moved by spin currents in the magnet continuously, which are appealing to imitate the biological neurotransmitters.

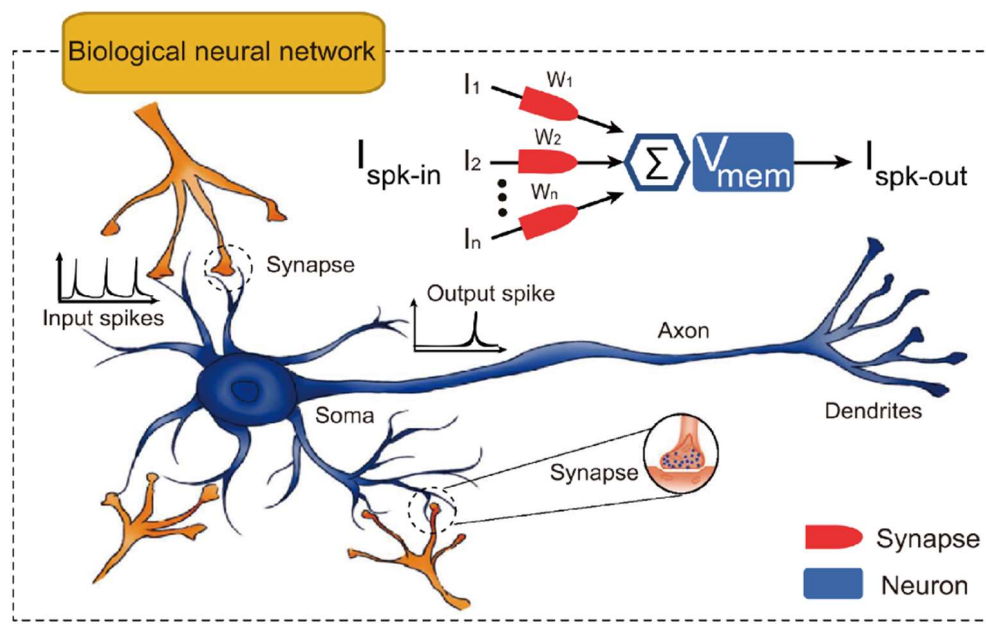

Fig.4. The skyrmionic neuromorphic computing devices. (a) Illustration of a biological nervous system composed of a neuron massively connected by a huge number of synapses and the simplified neural network model [66]. (Copyright 2017, IOP Publishing Group.)

In 2017, Huang et al. designed an artificial synapse device made of a nanotrack with multiple skyrmions, which could realize synaptic plasticity in the pre- and post-synapses (Fig. 5a) separated by a gated barrier at the center [67]. The main structure of the device is composed of FM layer/heavy metal (HM) hosting perpendicular magnetic anisotropy (PMA) and interfacial DMI, which help the transportation of skyrmions. Here, the number of skyrmions under the detection MTJ represents the synapse conduction ability (synaptic weight in neural networks), giving rise to the resistance signal of the device. During the learning phase, a bidirectional spin current induced by the SOT will drive skyrmions into (or out of) the post-synapse region to increase (or decrease) the synaptic weight, mimicking the potentiation/depression process of a biological synapse. By tuning the critical current density for skyrmion motion and the energy barrier in the nanotrack, both short-term plasticity (STP) and long-term potentiation (LTP) functionalities have been demonstrated for a spiking time-dependent plasticity (STDP) scheme, based on this device. Recently, Song et al. have experimentally demonstrated such an electrically-operating skyrmionic synapse device (Fig. 5b), using controlled current-induced creation, motion, deletion of skyrmions in ferrimagnetic multilayers [68].

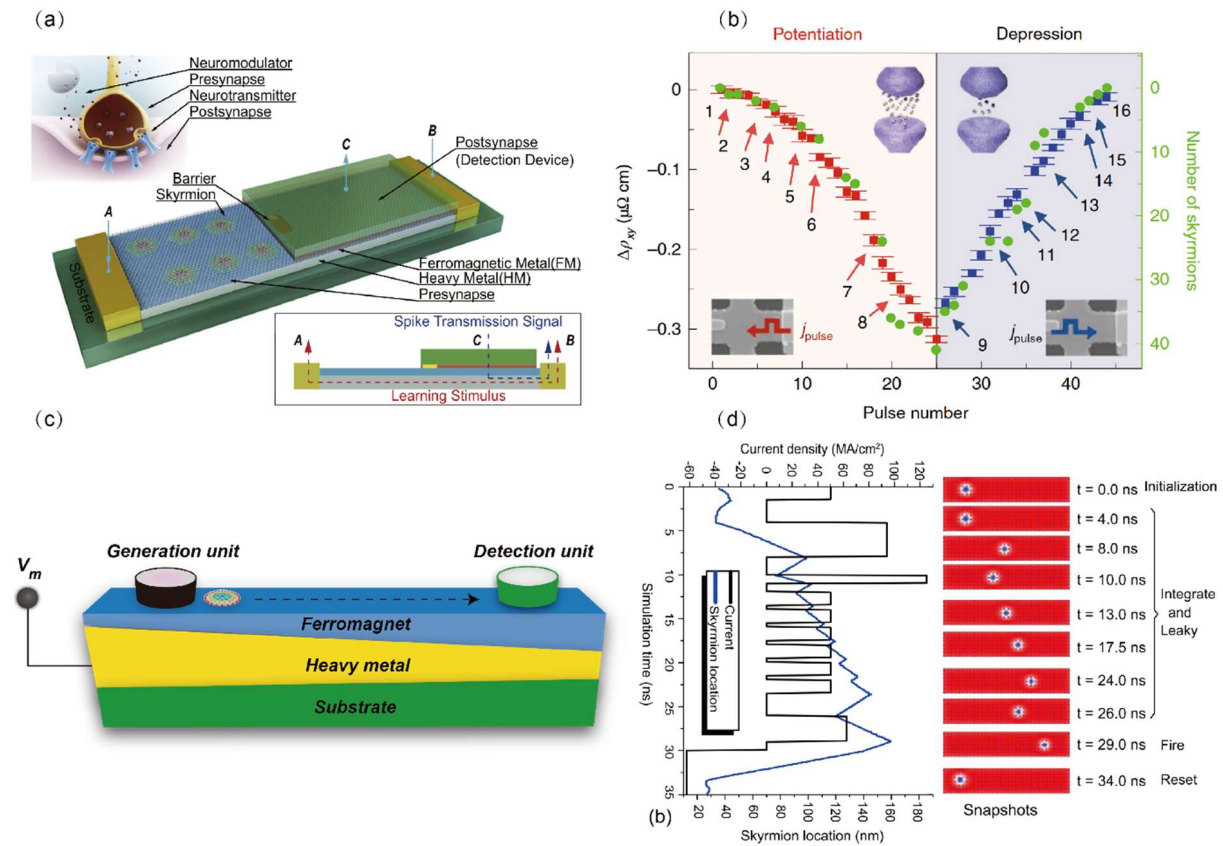

Fig.5. (a) Schematic of a biological synapse and the proposed skyrmionic synaptic device. To mimic a neuromodulator, a bidirectional learning stimulus flowing through the heavy metal, mimicking the potentiation/depression process of a biological synapse [67]. (Copyright 2017, IOP Publishing Group.) (b) The synaptic resistivity change and skyrmion number variation as a function of pulse number according to the potentiation/depression mode [68]. (Copyright 2020, Nature Publishing Group.) (c) Schematic of the proposed skyrmionic artificial neuron device, which hosts a linear increase of the PMA value along the nanotrack. (d) The skyrmion motion distance along the nanotrack following the LIF model as a function of spiking current pulse [66]. (Copyright 2017, IOP Publishing Group.)

## Skyrmionic neuron devices

The functional mechanism of a biological neuron is complicated, where an equilibrium membrane potential is maintained nonlinearly under the excitatory or inhibitory of the connected synaptic signals. The leaky-integrate-fire (LIF) model has been widely adopted for emulating a spiking neuron, described by [69,70],

$$\tau_{mem} \frac{dV_{mem}}{dt} = -(V - V_{rest}) + \sum_j \delta(t - t_j) w_j \tag{3}$$

When a neuron receives spikes, an all-or-nothing pulse called membrane potential in the neuron will accumulate with a leaky-integrate pattern. Once the potential reaches the threshold, the neuron will emit an output spike and then reset. Until now, less nanoscale devices were proposed to emulate neuron compared to synapse, with more complicated features than synapse, such as memristors device [71], domain wall motion (DWM) device [72], and insulator-to-metal (IMT) device [73].

Such a LIF model was behaviorally described by Li et al. through the use of the nonlinear skyrmion dynamics on a nanotrack (Fig. 5c), indicating the membrane potential of a biological neuron [66]. To obtain a linear PMA along the nanotrack, the primary component of the device is a thickness-gradient FM layer with a uniform HM layer. Although the difficulty of this design is to fabricate thickness gradient sample with stable skyrmion motion because of possible varying DMI and other magnetic properties, it has been already reported in the experiment with motion of chiral DW and skyrmion bubbles under special conditions [74]. The generation and detection units of the skyrmion are located on both sides of the nanotrack, where the pre-spike current force and the backward PMA force enable the skyrmion mobile along the nanotrack forth and back, competing with each other. Once the skyrmion reaches the detection area, the neuron "fires" an output spike, followed by the reset operation of the neuron with the skyrmion moving back to the initial location under a reset current (Fig. 5d). This device holds the promise to be integrated with emerging SOT memory, sharing the nanotrack to process and store the information bits together.

A similar prototype of skyrmionic neuron device was designed based on a wedge-shaped nanotrack with width varying [75], where competition forces lie between the driving current force and the nanotrack edge repulsive force. This scheme demands for untraditional method to confine the skyrmion artificially, which may be realized by more advanced nanofabrication technology. Another study presents an application of a fixed magnetic skyrmion in an MTJ with oscillation of a few Gigahertz for resonate and fire type neuron, different from the LIF model [76]. Such design has potential in coupled nanomagnetic oscillator based memory arrays. In addition, He et al. [77] proposed a skyrmionic neuron cluster (SNC) to approximate non-linear soft-limiting neuron activation functions, such as the sigmoid function by adaption to current spiking neuron networks (SNN) neural model. In summary, the skyrmionic neuron devices could mimic the biological neuron by their controllable mobility with shape-gradient or anisotropy-induced device, which is similar to the DMW devices [78,79]. Nevertheless, skyrmion is easier to escape the impurity's pinning force [80] thus the device is more robust, and the skyrmionic device could have more diverse shapes not only the nanotrack. These studies of skyrmionic artificial neuron devices would boost the dense, fast and energy efficient neuromorphic computing design in the future.

### 3.3 Skyrmionic reservoir computing devices

The concept of reservoir computing (RC) roots in special arithmetic features of recurrent neural network (RNN), mainly derived from echo state networks (ESNs) [81] and liquid state machines (LSMs) [82]. To date, RC has already exceeded the limit of the conceptual RNN, acting as a temporal 'kernel' to conduct the data pre-processing [83] or a remedy for the large demand of training in certain spatio-temporal applications [84]. Uniquely, RC lies in a randomly created medium called reservoir, which could convert the input into spatiotemporal patterns in a high-dimensional feature space (Fig. 6a). This feature makes it especially applicable to solve the temporal/sequential problems even without training the weights within the input and reservoir layers. Various technologies simplifying the internal topologies of the reservoir have demonstrated the potential applications of RC [85], such as photonics, RRAM, and atomic switches. With the dominance over fast information processing speed and low learning cost, plenty of spintronic reservoir devices have also been proposed to implement RC, including STNO[86] and spin wave [87].

Alternatively, skyrmion is a good fit for RC as a spiking neural node in the reservoir due to its internal degrees of freedom with complex interactions with currents and magnons. Pychynenko et al. first theoretically explored the pinning-induced nonlinear voltage characteristics of skyrmion for RC design (Fig. 6b), which originates from the interaction of spin torques and the magnetoresistance in a two-terminal device with skyrmion embedded in magnetic films (Fig. 6c) [88]. The voltage between the two contacts will spur the skyrmion movement or deformation, according to the shape, type, relative size and position of the skyrmion. Notably, the random thermal fluctuations may even enhance the nonlinear signals, where the negative effect of the temperature on conventional device, e.g. racetrack memory or logic gate, is converted into an advantage on the device performance. In additional to single skyrmion, the more complex magnetic texture of skyrmion fabrics could also generate a non-linear current flow, responding to different voltages [89], which has more tunability by optimizing the material. Compared to other physical solutions of RC, skyrmion could offer a rich array of topology-dependent static and dynamic properties to support a reservoir for spatial-temporal event, such as the skyrmion-skyrmion interaction, thermal diffusion, and breathing mode. In 2019, Jiang et al. implemented one single magnetic skyrmion memristor with its nonlinear responses directly acting as an output to process classification task, well demonstrating skyrmion-based RC system on digital image classification [90].

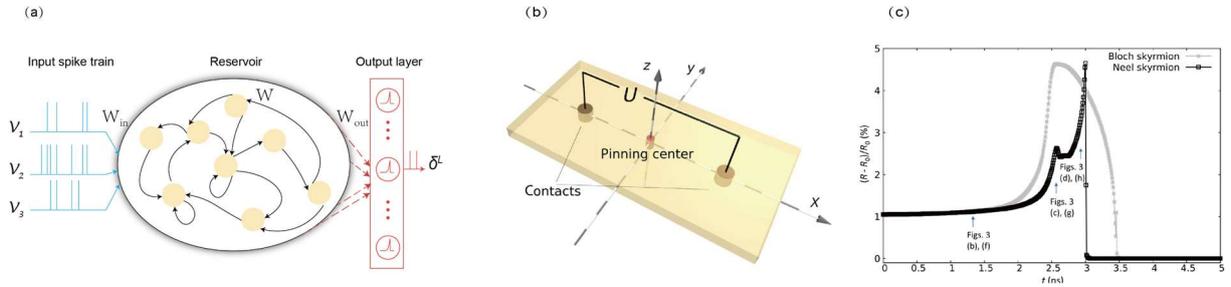

fig.6. (a) The structure of a RC network: W means the weights in reservoir layer. (b) Schematic of the skyrmionic RC device. A voltage U is applied between two contacts of the FM thin film, where skyrmion is depinned and driven by the voltage. A potential pinning center is located in the center of the film. (c) Relative resistance dependence (R-R0)/R0 on time for Neel-type and Bloch-type skyrmions at a fixed voltage [88]. (Copyright 2018, American Physical Society.)

### 3.4 Skyrmionic stochastic computing devices

Different from mainstream computing technologies using numerical values encoded as conventional binary format, stochastic computing (SC) processes random bit streams continuously. In this content, SC can be implemented much more simply by bit-wise operations to replace complicated digital computations at low cost and with error tolerance. For instance, a bit-stream containing 25% 1s and 75% 0s indicates the value p = 0.25, reflecting a probability of observing a 1 at an arbitrary bit position of the stream, without fixing the length and the structure [91]. For example, a multiplication of two input bit-streams p1 and p2 can be computed by a single AND gate, where the output shows p1 × p2 (Fig. 7a) [92]. If a single bit is changed in a long bit-stream, the output stochastic number only alters a little, showing good error-tolerance. However, it is important to make the two bit-streams suitably uncorrelated or independent in SC. An extreme case is that two identical bit-streams will lead to a wrong output, far from the correct one (Fig. 7b). Therefore, the great demand for SC is to reshuffle the input signals into uncorrelated streams before logic operations. In conventional semiconductor circuits, a pseudorandom number generator combining with a shift register is used to generate or to reshuffle signals, resulting in high complexity, high hardware cost and low energy efficiency.

The thermal-induced skyrmion random motion has recently been demonstrated in theory and in experiment [93], which provides a promising route to use skyrmion as basic elements in SC design. In this context, a skyrmionic reshuffle was proposed by Pinna et al. in 2018 [92], where inconvenient correlations can be washed out effectively (Fig. 7c). The skyrmionic reshuffle device consists of two circular chambers with input-output conduit tracks ushering skyrmions in and out. The net drift of skyrmion is forced by a static current flowing across the entire structure. The up-and-down (binary) states of the bit-stream are divided into two pulses to select which chamber would inject the skyrmion at a constant rate according to the input's state (Fig. 7d). To ensure the p-value of the output from the chamber is identical to that of the input signal, whenever an outgoing skyrmion is read-out from the up (down)-chamber, we switch the outgoing signal into an up (down) state. Its p-value will be identical to that of the input signal as long as the relative skyrmion numbers in the chamber are conserved. This skyrmion reshuffler could help tackle the above long-standing problem of SC. Followed by this innovative concept, thermally excited skyrmion diffusion was experimentally observed and used to construct a reshuffler device by Zázvorka et al. in 2019 [94] (Fig. 7e). The proof-of-concept reshuffler device successfully demonstrates that the thermal induced skyrmion dynamics could create uncorrelated signal streams. It should be noted that all-skyrmion-based SC system may not need the reshuffler device because its motion has intrinsic stochasticity, which is an important advantage for SC and needs to be verified with additional research in the future.

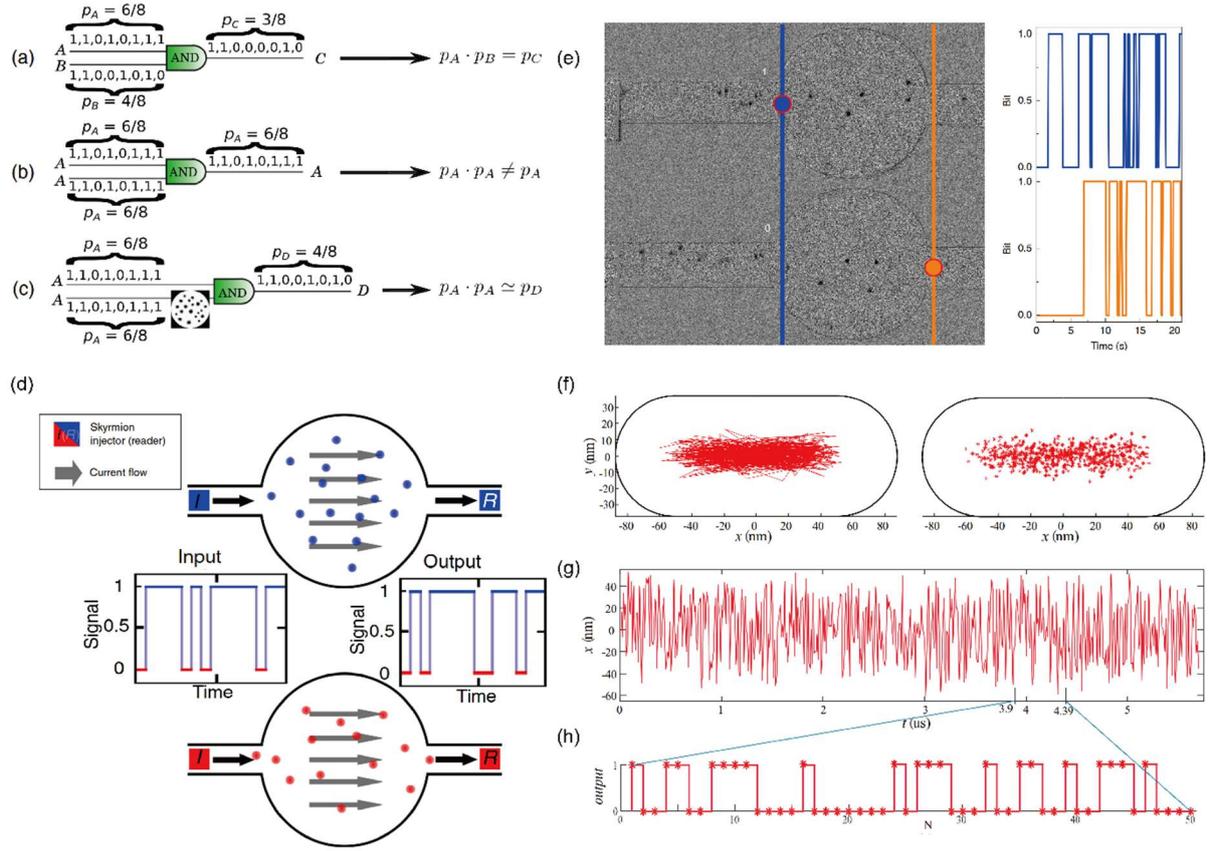

Fig.7. The schematic diagram of stochastic AND gate used as multiplication operation: (a) The p-value of the output is the exact product of input signals. (b) The output result is not the expected value because there are correlations in the two input signals. (c) The output value is an approximate result of the exact one if the input signals pass through a reshuffler. (d) The skyrmionic reshuffler device consisting of two magnetic chambers. The input signals are injected into the selected chamber depending on the state of the input signal. The output order is reshuffled from the input signals because of the thermal diffusion of skyrmions [92]. (Copyright 2018, American Physical Society.) (e) Experimental demonstration of the skyrmion reshuffler device. Reshuffler operation with skyrmion nucleation by a direct current. The input signal is constructed as a time frame in which the skyrmion crosses the blue threshold line, while the output is determined on the skyrmion crossing the orange line. The right figure is the corresponding input (blue) and reshuffled output signals (orange) [94]. (Copyright 2019, Nature Publishing Group.) (f) Random bit sequence generated by skyrmion Brownian motion. The motion trajectory of the skyrmion in the chamber. (g) A random bit sequence obtained by detecting the relative position of skyrmion in the x-axis. (h) Outputs of selected 50 bits [95] .(Copyright 2020, IEEE Publishing Group.)

How to efficiently build a true random number generator (TRNG) is another important issue in SC. Conventional CMOS-based solutions, e.g. oscillator sampling or directly noise amplifying TRNGs usually consume a large hardware area and high power consumption, offsetting the main advantages of SC [96]. To exploit the intrinsic random behaviors of skyrmion, Yao et al. designed a skyrmionic TRNG utilizing the continuous thermal induced skyrmion Brownian motion in a confined geometry [95], which is a rectangular region with two semicircular regions (Fig. 7f). Thanks to the thermal fluctuations of skyrmion dynamics, the relative position of skyrmion inside the chamber assigns the output bits through the differential voltage of the two detection MTJs. The readout sequence is proven to be non-repeatable and unpredictable (Fig. 7g, h), which is evaluated by the National Institute of Standards and Technology (NIST) suite. Furthermore, a probability-adjustable skyrmionic TRNG is proposed, where a desired ratio of "0" and "1" can be obtained simply by introducing an anisotropy gradient through the VCMA effect.

Except these above-mentioned types of skyrmionic unconventional computing, there are also some very promising areas not detailedly introduced due to the limited space. Reversible computation could be implemented by the abundant physics of skyrmion, which could reduce the energy consumption and improve the computing efficiency [97]. Quantum computing may also combine with the skyrmionic studies in the future [98,99].

Table.1 Features of skyrmionic unconventional computing devices.

| Device | | Skyrmionic dynamics/Behaviros | Highlight | References |
|---|---|---|---|---|
| Microwave signal sources | | Skyrmion excited into the circle-round a disk by current | Stable small size, ultra-high-density data encoding | 44 |
| Microwave detectors and energy harvesting | | Skyrmion breathing mode in a MTJ by current | Sensitive, zero bias current, low input microwave power | 45 |
| Synaptic device | Simulation | Multiple skyrmions' motion in a track holding an energy barrier | Low power consumption, most similar biological principles | 67 |
| | Experiment | Measured Hall resistivity change due to injected current pulses, good linear weight | Experimental demonstration of the device at room temperature | 68 |
| Neuron device | | Skyrmion motion in a thickness-gradient nanotrack with PMA force | Small size, low power consumption and practical in experiments | 66 |
| Neuron device | | Skyrmion motion in a wedge-shaped nanotrack with edge force | Small size, low power consumption | 75 |
| Reservoir computing device | | One skyrmion motion/deformation in a nanotrack with a pinning center. | Small size, high Energy efficiency and tunability | 88 |
| | | Multi-domain skyrmion fabrics | More tunability and practical significance | 89 |
| Stochastic computing reshuffle device | Simulation | Inject skyrmions into the two chambers and reshuffling them randomly | Small driving currents, thermal noise makes skyrmion optimal candidates | 92 |
| | Experiment | Thermally excited skyrmion diffusion observed by MOKE | Proof-of-concept reshuffler device Reveals the possibility of Problistic computing | 94 |
| Stochastic computing TRNG device | | Thermal-induced skyrmion Brownian motion in a confined geometry | Non-repeatable and unpredictable TRNG | 95 |
| Reversible computation device | | The designed skyrmion motion with SkHE in the confined track | Reduced energy consumption and improved computing efficiency | 97 |

## 4    Application of skyrmionic unconventional computing: a case study

This section presents a case study of skyrmionic neuromorphic computing through the artificial neuron network (ANN) platform utilizing the room-temperature experimental results of skyrmion synapse mentioned in section 3.1 [68]. The ANN is a tri-layer structure (Fig. 8a) with only one hidden layer composed of 100 neurons to implement the pattern recognition task based on the handwritten pattern of Modified National Institute of Standards and Technology (MNIST) dataset, where the input layer is divided into 784 neurons. The skyrmion-based synapse devices connecting the pre- and post- neurons in the circuit diagram with varying conductance values and a normal distribution (Fig. 8b). The results demonstrated that the accuracy of such a design can achieve ~89%, comparable to that of software-based ANN of ~93%. We believe that further improvement of skyrmion in terms of size, material engineering, device structure and manipulation schemes would pave the pathway towards a scalable family of skyrmion-based unconventional computing.

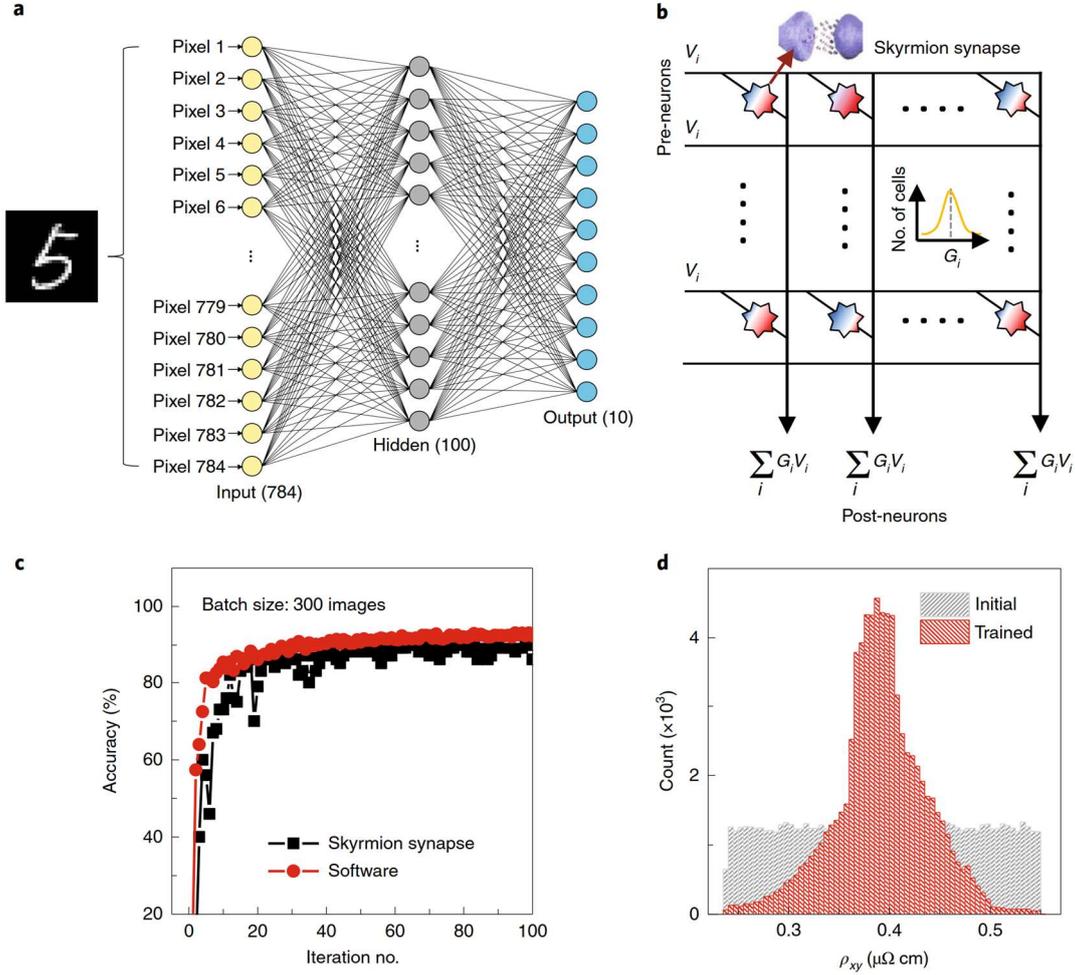

Fig.8. The pattern recognition through the skyrmion-based synapse device. (a) The ANN structure consists of a input layer, a hidden layer and a output layer. (b) Schematic of the circuit diagram comprising skyrmion-based synapses. (c) The pattern recognition accuracy as a function of training iteration. (d) Distribution of the skyrmion synapse resistivity before and after training [68]. (Copyright 2020, Nature Publishing Group.)

# 5 Challenges and Perspectives

To exploit the full potential of skyrmion devices for unconventional computing, this section presents the challenges and highlights some potential future studies (Fig. 9a) from an application perspective.

## 5.1 Material optimization and experimental demonstration

In skyrmion-related materials, recent experiments have focused on two essential directions: (a) how to obtain large DMI and adaptive anisotropy $K_{\text{eff}}$, and (b) fast and energy-efficient electric filed/current-induced skyrmion motion. A typical structure hosting skyrmions is HM material (e.g. Pt, W, Ta, Ir) adjacent to FM layer (e.g. Co, CoFeB), which could provide tunable anisotropy and DMI at the interface by tuning the thickness of the layer. In general, a smaller $K_{\text{eff}}$ and a proper DMI are preferable to obtain more stable skyrmions at room temperature. However, too weak $K_{\text{eff}}$ would cause deformation of skyrmions, limiting the skyrmion velocity and stability. In addition, it has been found that a smaller skyrmion would have a slower velocity under the certain driving current density [100]. Therefore, a better understanding of how to decorrelate these magnetic parameters by tuning the material or device structure is necessary to help control the skyrmion size and its dynamics. Moreover, the dipolar interactions in FM films would block the stablity of ultrasmall skyrmions. How to trade off the skyrmion stablity, device size, speed, and energy comsuption would be a significant but challenging topic in unconventional computing device design.

Beyond ferromagnets, new material systems, such as compenasted materials, ferrimagnet and antiferromagnet (AFM), have been employed for skyrmions. Owing to the reduced saturation magnetization $M$s in these materials, the spin torque could be more effeicienly applied to the magnetization, thus improving the mobility of skyrmion. Another advantage is that SkHE could be significantly eliminated due to the zero net topological charge [101], which is also applicable for skyrmions in synthetic antiferromagnetic (SAF) structure (e.g. Pt/Co/Ru). A recent experiment has confirmed that room-temperature skyrmion bubbles can be moved by SOT current with negligible SkHE in SAF systems [102]. Furthermore, 2D van der Waals (vdW) materials have draw considerable attention for its unique SOC origins. It was reported that Néel-type skyrmion crystals as well as isoalted Neel-type skyrmions can be found in a vdW ferromagnet $Fe_3GeTe_2$ (FGT) flake [103], where skyrmions can be driven into motion by spin currents. Wu et al. demonstrated that WTe2/FGT vdW heterostructure could induce interficial DMI and Néel-type skyrmions[104]. In addition, skyrmions in frustrated magnet [105] and artificial skyrmion crystal [106] could also be stablized in the absense of DMI. However, these exsiting material systems all require further proof of compatibility with CMOS integration technologies and remain to be explored in more internal degrees of freedom by device designers, which could provide alternative routes for future skyrmionic unconventional computing.

## 5.2 Device-circuit codesign and hybrid simulation framework

For current skyrmionic unconventional computing, design space exploration was limited at the device level to demonstrate the functionality and analyze the impact of various parameters (e.g. PMA, DMI, and device dimension) on the required device performance. However, there is a gap between the device and circuit design. The necessity of carrying out device-circuit codesign is highly important and will facilitate the designer to meet the system-level optimization goals and design requirements. The researchers in [107] proposed an all-spin design demo with skyrmionic neurons and synapses, exploring a systematic device-circuit-architecture co-design for digit recognition. The results show that the skyrmionic design can potentially improve two orders of magnitude in energy consumption in comparison to a CMOS solution at 45 nm technology node. This kind of study would greatly accelerate the research and development of skyrmion-electronics. Meanwhile, only a few studies have involved the monolithic integration of skyrmion and silicon transistor. This leaves much space for future explorations of circuit parameters (e.g. bit-cell, layout and routing), which may be quite different to traditional circuit studies. Furthermore, the peripheral circuitry (e.g. the generating, detecting and shifting circuits of skyrmion) will largely determine the extent to the latency and energy consumption. Therefore, device-circuit codesign could accelerate device optimization and explore design space. Furthermore, hybrid simulation framework/tool is essential and allows system architects to analyze the issues and the impact of skyrmionic unconventional computing in terms of performance, energy and reliability characteristics before its integration into electronic systems.

## 5.3 The outlook of skyrmion-based unconventional computing

In order to present the most necessary targets of future skyrmion-based unconventional computing, we describe some attributes as below (Fig. 9b):

(1) Density. It refers to how many devices could be integrated on one chip, and has always been one of the core parameters for high-performance computing. It has been theoretically reported that the skyrmion size could reach 1 nm. And a recent experimental study has demonstrated that skyrmions in SAFs could be smaller than 10 nm at room temperature [108], offering a great advantage in density compared to other emerging technologies. Nevertheless, the peripheral circuitry and process technology should be accordingly optimized to fully exploit the size advantage of skyrmions.

(2) Nonvolatility. It means how long the information could remain stable after power-off. Skyrmion is magnetic object and thus physically nonvolatile, however, the stability at room temperature needs more in-depth study, withstanding to noise and thermal fluctuations.

(3) Power consumption. It includes the energy consumption for skyrmion generation, detection and manipulation. A variety of mechanisms and approaches have been proposed to achieve low-power operation of skyrmionic devices.

(4) Endurance. It refers to how many times the device can be operated, which is particularly important for computing applications with frequent access. Fortunately, it is envisioned that skyrmion (spin state) could be created/deleted repeatedly without damaging the device.

(5) Maturity. Maturity is extremely important for practical applications. Over the last decade, the physics society has made great advances on skyrmionic devices, but its potential at the circuit- and architecture-level needs more investigation.

(6) Latency. It mainly arises from the package of multiple information bits in one device as well as the peripheral circuitry, which generally can be optimized with various methods.

(7) Variability. It represents the device variation in the material engineering and fabrication process. Fortunately, variability can be well tolerated or even offer an advantage in accuracy-insensitive unconventional computing.

(8) Analog capability. It generally refers to the capability to display continuous status modulation determined by the spatial-temporal flexibility of the device. This happens to be the superiority of skyrmionic device, as has already been demonstrated in the skyrmionic neuromorphic computing devices and reservoir computing devices.

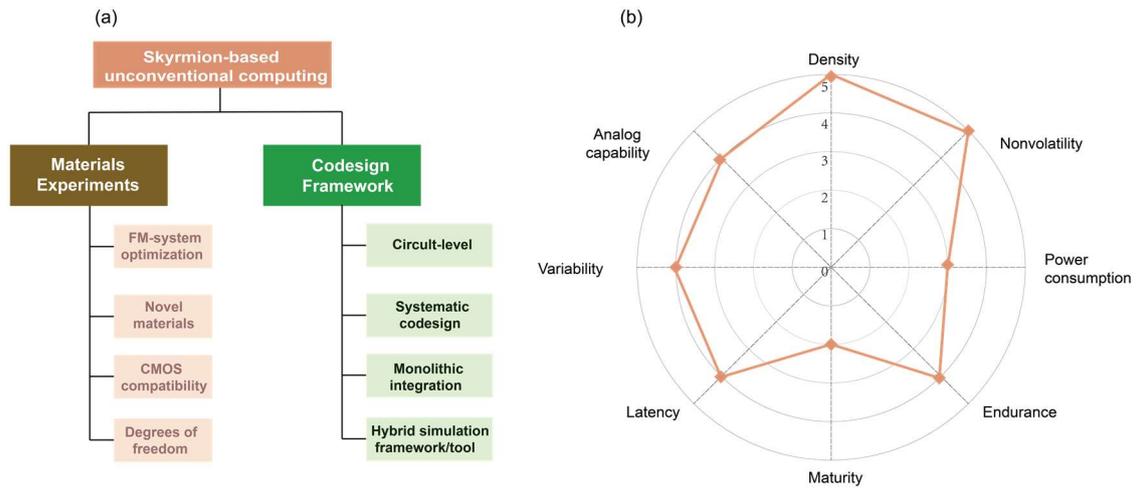

Fig.9. (a) An overview of the current challenges of skyrmion-based unconventional computing. (b) The spider chart presenting the most necessary targets of future unconventional computing and the current status of skyrmionic devices.

## 6 Conclusion

The discovery of skyrmion in magnetic materials and the physical advances in the last decade have paved the way for bringing skyrmion to applications. In particular, the device studies and the intrinsic dynamic behaviors highlight the positive prospects of skyrmion for unconventional computing, which would enable potential alternative/complementary of CMOS in post-Moore era. Moreover, recent experimental demonstrations in skyrmionic neuromorphic computing and stochastic computing devices show the practical possibility. From an application perspective, the development of device-circuit codesign framework and hybrid simulation tool, which could be an essential research direction, would rapidly accelerate the skyrmionic device optimization and its exploration in unconventional computing systems. This paper aims to enable readers to quickly build up a picture of skyrmionic unconventional computing, simultaneously to inspire more effort to this interdisciplinary research area.

# Funding

S.L. and W.K. gratefully acknowledge the Beijing Natural Science Foundation (Grants No. 4202043), Beijing Nova Program (Z201100006820042) from Beijing Municipal Science and Technology Commission, National Natural Science Foundation of China (Grants No. 61871008). T.N. was supported by the National Key R&D Program of China (2018YFB0407602), the National Natural Science Foundation of China (61774013). W.Z. thank the National Natural Science Foundation of China (Grant No. 61627813), the International Collaboration Project B16001. X.Z. was supported by the Guangdong Basic and Applied Basic Research Foundation (Grant No. 2019A1515110713), and the Presidential Postdoctoral Fellowship of The Chinese University of Hong Kong, Shenzhen (CUHKSZ). Y.Z. acknowledges the support by Guangdong Special Support Project (2019BT02X030), Shenzhen Peacock Group Plan (KQTD20180413181702403), Pearl River Recruitment Program of Talents (2017GC010293) and National Natural Science Foundation of China (11974298, 61961136006).

# Competing interests

The authors declare no competing interests.

# Data availability statements

The data that support the findings of this study are available from the corresponding author upon reasonable request.